\PassOptionsToPackage{unicode}{hyperref}
\PassOptionsToPackage{hyphens}{url}
\PassOptionsToPackage{dvipsnames,svgnames,x11names}{xcolor}
\documentclass[
  11pt,
  number,
  preprint]{elsarticle}

\usepackage{amsmath,amssymb}
\usepackage{iftex}
\ifPDFTeX
  \usepackage[T1]{fontenc}
  \usepackage[utf8]{inputenc}
  \usepackage{textcomp} 
\else 
  \usepackage{unicode-math}
  \defaultfontfeatures{Scale=MatchLowercase}
  \defaultfontfeatures[\rmfamily]{Ligatures=TeX,Scale=1}
\fi
\usepackage{lmodern}
\ifPDFTeX\else  
\fi
\IfFileExists{upquote.sty}{\usepackage{upquote}}{}
\IfFileExists{microtype.sty}{
  \usepackage[]{microtype}
  \UseMicrotypeSet[protrusion]{basicmath} 
}{}
\makeatletter
\@ifundefined{KOMAClassName}{
  \IfFileExists{parskip.sty}{%
    \usepackage{parskip}
  }{
    \setlength{\parindent}{0pt}
    \setlength{\parskip}{6pt plus 2pt minus 1pt}}
}{
  \KOMAoptions{parskip=half}}
\makeatother
\usepackage{xcolor}
\usepackage[top=30mm,bottom=30mm,left=30mm,right=30mm]{geometry}
\makeatletter
\ifx\paragraph\undefined\else
  \let\oldparagraph\paragraph
  \renewcommand{\paragraph}{
    \@ifstar
      \xxxParagraphStar
      \xxxParagraphNoStar
  }
  \newcommand{\xxxParagraphStar}[1]{\oldparagraph*{#1}\mbox{}}
  \newcommand{\xxxParagraphNoStar}[1]{\oldparagraph{#1}\mbox{}}
\fi
\ifx\subparagraph\undefined\else
  \let\oldsubparagraph\subparagraph
  \renewcommand{\subparagraph}{
    \@ifstar
      \xxxSubParagraphStar
      \xxxSubParagraphNoStar
  }
  \newcommand{\xxxSubParagraphStar}[1]{\oldsubparagraph*{#1}\mbox{}}
  \newcommand{\xxxSubParagraphNoStar}[1]{\oldsubparagraph{#1}\mbox{}}
\fi
\makeatother

\providecommand{\tightlist}{%
  \setlength{\itemsep}{0pt}\setlength{\parskip}{0pt}}\usepackage{longtable,booktabs,array}
\usepackage{calc} 
\usepackage{etoolbox}
\makeatletter
\patchcmd\longtable{\par}{\if@noskipsec\mbox{}\fi\par}{}{}
\makeatother
\IfFileExists{footnotehyper.sty}{\usepackage{footnotehyper}}{\usepackage{footnote}}
\makesavenoteenv{longtable}
\usepackage{graphicx}
\makeatletter
\newsavebox\pandoc@box
\newcommand*\pandocbounded[1]{
  \sbox\pandoc@box{#1}%
  \Gscale@div\@tempa{\textheight}{\dimexpr\ht\pandoc@box+\dp\pandoc@box\relax}%
  \Gscale@div\@tempb{\linewidth}{\wd\pandoc@box}%
  \ifdim\@tempb\p@<\@tempa\p@\let\@tempa\@tempb\fi
  \ifdim\@tempa\p@<\p@\scalebox{\@tempa}{\usebox\pandoc@box}%
  \else\usebox{\pandoc@box}%
  \fi%
}
\def\fps@figure{htbp}
\makeatother

\usepackage{booktabs}
\usepackage{longtable}
\usepackage{array}
\usepackage{multirow}
\usepackage{wrapfig}
\usepackage{float}
\usepackage{colortbl}
\usepackage{pdflscape}
\usepackage{tabu}
\usepackage{threeparttable}
\usepackage{threeparttablex}
\usepackage[normalem]{ulem}
\usepackage{makecell}
\usepackage{xcolor}
\usepackage{placeins}
\makeatletter
\@ifpackageloaded{caption}{}{\usepackage{caption}}
\AtBeginDocument{%
\ifdefined\contentsname
  \renewcommand*\contentsname{Table of contents}
\else
  \newcommand\contentsname{Table of contents}
\fi
\ifdefined\listfigurename
  \renewcommand*\listfigurename{List of Figures}
\else
  \newcommand\listfigurename{List of Figures}
\fi
\ifdefined\listtablename
  \renewcommand*\listtablename{List of Tables}
\else
  \newcommand\listtablename{List of Tables}
\fi
\ifdefined\figurename
  \renewcommand*\figurename{Figure}
\else
  \newcommand\figurename{Figure}
\fi
\ifdefined\tablename
  \renewcommand*\tablename{Table}
\else
  \newcommand\tablename{Table}
\fi
}
\@ifpackageloaded{float}{}{\usepackage{float}}
\floatstyle{ruled}
\@ifundefined{c@chapter}{\newfloat{codelisting}{h}{lop}}{\newfloat{codelisting}{h}{lop}[chapter]}
\floatname{codelisting}{Listing}

\makeatother
\makeatletter
\@ifpackageloaded{caption}{}{\usepackage{caption}}
\@ifpackageloaded{subcaption}{}{\usepackage{subcaption}}
\makeatother
\journal{arXiv}

\usepackage[]{natbib}
\usepackage{bookmark}

\IfFileExists{xurl.sty}{\usepackage{xurl}}{} 
\hypersetup{
  pdftitle={MitoFREQ: A Novel Approach for Mitogenome Frequency Estimation from Top-level Haplogroups and Single Nucleotide Variants},
  pdfauthor={Mikkel Meyer Andersen; Nicole Huber; Kimberly S Andreaggi; Tóra Oluffa Stenberg Olsen; Walther Parson; Charla Marshall},
  pdfkeywords={Evidential weight, Forensic genetics, mtDNA, Likelihood
ratio, Match probability},
  colorlinks=true,
  linkcolor={blue},
  filecolor={Maroon},
  citecolor={Blue},
  urlcolor={Blue},
  pdfcreator={LaTeX via pandoc}}

\begin{document}

\begin{frontmatter}
\title{MitoFREQ: A Novel Approach for Mitogenome Frequency Estimation
from Top-level Haplogroups and Single Nucleotide Variants}
\author[1,2]{Mikkel Meyer Andersen%
\corref{cor1}%
}
 \ead{mikl@math.aau.dk} 
\author[3]{Nicole Huber%
}

\author[4,5]{Kimberly S Andreaggi%
}

\author[1]{Tóra Oluffa Stenberg Olsen%
}

\author[3,6]{Walther Parson%
}

\author[4,6]{Charla Marshall%
}

\affiliation[1]{organization={Aalborg University, Department of
Mathematical
Sciences},city={Aalborg},country={Denmark},countrysep={,},postcodesep={}}
\affiliation[2]{organization={University of Copenhagen, Section of
Forensic Genetics, Department of Forensic
Medicine},city={Copenhagen},country={Denmark},countrysep={,},postcodesep={}}
\affiliation[3]{organization={Medical University of Innsbruck, Institute
of Legal
Medicine},city={Innsbruck},country={Austria},countrysep={,},postcodesep={}}
\affiliation[4]{organization={Armed Forces Medical Examiner
System, Armed Forces DNA Identification Laboratory},city={Dover,
Delaware},country={US},countrysep={,},postcodesep={}}
\affiliation[5]{organization={SNA International},city={Alexandria,
Virginia},country={US},countrysep={,},postcodesep={}}
\affiliation[6]{organization={The Pennsylvania State
University, Forensic Science Program},city={University
Park},country={US},countrysep={,},postcodesep={}}

\cortext[cor1]{Corresponding author}

\begin{abstract}
Forensic lineage markers pose a challenge in forensic genetics as their
evidential value can be difficult to quantify. Lineage marker population
frequencies can serve as one way to express evidential value. However,
for some markers, e.g., high-quality whole mitochondrial DNA genome
sequences (mitogenomes), population data remain limited. In this paper,
we offer a new method, MitoFREQ, for estimating the population
frequencies of mitogenomes. MitoFREQ uses the mitogenome resources
HelixMTdb and gnomAD, harbouring information from 195,983 and 56,406
mitogenomes, respectively. Neither HelixMTdb nor gnomAD can be queried
directly for individual mitogenome frequencies, but offers single
nucleotide variant (SNV) allele frequencies for each of 30 ``top-level''
haplogroups (TLHG), which mainly correspond to the first letter of major
mitochondrial DNA (mtDNA) haplogroups (e.g., A, B, C, D, E, etc.) except
for the L0, L1, L2, L3, L4-6, HV, and R/B haplogroups. We propose using
the HelixMTdb and gnomAD resources by classifying a given mitogenome
within the TLHG scheme and subsequently using the frequency of its
rarest SNV within that TLHG weighted by the TLHG frequency. We show that
this method is guaranteed to provide a higher population frequency
estimate than if a refined haplogroup and its SNV frequencies were used.
Further, we show that top-level haplogrouping can be achieved by using
only 227 specific positions for 99.9\% of the tested mitogenomes,
potentially making the method available for low-quality samples. The
method was tested on two types of datasets: high-quality forensic
reference datasets and a diverse collection of scrutinised mitogenomes
from GenBank. This dual evaluation demonstrated that the approach is
robust across both curated forensic data and broader population-level
sequences. This method produced likelihood ratios in the range of
100-100,000, demonstrating its potential to strengthen the statistical
evaluation of forensic mtDNA evidence. We have developed an open-source
R package \texttt{mitofreq} that implements our method, including a
Shiny app where custom TLHG frequencies can be supplied.
\end{abstract}

\end{frontmatter}

\section{Introduction}\label{introduction}

Forensic lineage markers pose a challenge in forensic genetics as their
evidential value can be difficult to quantify \citep{Andersen2021}. The
evidential value can be expressed as how many other individuals could
leave a trace that would yield the same trace profile
\citep{Andersen2017Y, Andersen2018mtDNA}. The problem for lineage
markers is that closely related individuals have a higher probability of
matching than remotely related individuals, and this is true for
individuals also further apart than a few generations and where the
close relationship may not be recognised
\citep{Andersen2021, Andersen2017Y, Andersen2018mtDNA}.

If the lineage marker DNA profiles have a sufficiently low mutation
probability, the distribution may be the same in the entire population
and population frequencies may serve as match probabilities
\citep{Andersen2021}.

This approach is useful when a matching profile is observed in the
database. For a relevant database of size \(n\) and a lineage marker DNA
profile, \(q\), that has been observed \(k_q\) times in the relevant
database, then if \(k_q \geq 1\), the binomial estimator \(k_q / n\) is
a good estimator of the population frequency (especially if
\(k_q > 1\)). The problem is when there is no match in the database
resulting in \(k_q = 0\), as is happening in the majority of the cases
\citep{Andersen2021}. There are various existing methods to deal with a
``no match'' situation (\(k_q = 0\)), including augmenting the database,
upper confidence limits (e.g., by the Clopper-Pearson formula),
Brenner's \(\kappa\) estimator \citep{Brenner2010}, and Cereda's
estimator based on the Generalized-Good-Turing (CGGT) estimator
\citep{Good1953, Cereda2016}. Augmenting the database works by adding
one or two copies of the profile \(q\) to the database and using the
binomial estimator on the extended database \citep{Andersen2023}
yielding the frequency estimators \(1 / (n + 1)\) and \(2 / (n + 2)\),
respectively. The upper confidence limit works by using the binomial
distribution and asking what is the highest value of the population
frequency that with some confidence (typically 95\%) can result in not
observing any copies in a sample of size \(n\). All these binomial-based
estimators (\(k_q / n\), \(1 / (n + 1)\), \(2 / (n + 2)\), and upper
confidence limit/Clopper-Pearson) assume that all possible profiles have
been observed. Thus, they leave no probability left to be assigned to
profiles not yet observed. Since new profiles frequently have no match
in the database, this assumption is violated. This problem is addressed
by both Brenner's \(\kappa\) and the CGGT estimators, that, in simple
terms, work by multiplying the augmented database frequency
\(1 / (n + 1)\) by deflation factors to incorporate the fact that there
are profiles we still have not seen. Except for the simple augmented
database estimators \((k_q + 1)/(n+1)\) and \((k_q + 2)/(n+2)\),
Brenner's \(\kappa\) and the CGGT estimators are only valid for
singletons (upper confidence limit/Clopper-Pearson can be used for
others, although it results in a rather strange concept). And the
augmented database estimators still suffer from the problem of assuming
that all possible profiles are known.

Thus, there is a need for a method that works in the same way regardless
of previous database occurrences of the profile of interest.

Population data remain limited for whole mitochondrial DNA genome
sequences (mitogenomes): The EMPOP database \citep{Parson2007} currently
only contains 10,648 mitogenomes, representing 0.00013\% of the human
population of 8.2 billion. Alternative mitogenome databases developed by
the biomedical and population genetics fields, HelixMTdb
\citep{Bolze2019} and gnomAD \citep{Laricchia2022, Chen2023}, have
larger sample sizes (195,983 and 56,406, respectively). Although they
cannot be queried by entire sequence strings (or called profiles) as
currently done in forensic profile searching, both HelixMTdb and gnomAD
contain searchable haplogroup (HG) and single nucleotide variant (SNV)
frequency data. In order to make use of these larger mitogenome
datasets, an alternative approach to searching the database is
necessary.

In this paper, we describe a simple method for estimating population
frequencies of mitogenomes. We use only the ``top-level'' haplogroup
(TLHG) and one SNV with frequencies from publicly available databases.
Thus, the overall mutation rate is low and we can use relatively few
markers which makes the method robust and useful for low-quality samples
with partial profiles (i.e., incomplete interpretation ranges with less
than 16,569 bases reported).

Improving mitogenome population frequency estimates will benefit many
forensic contexts that rely on mitochondrial DNA evidence
interpretation. These include cases involving hair shafts, historical
remains, or very distant maternal relatives \citep{King2014} - those
beyond the limit of detecting nuclear genetic relationships (nine or
more degrees of separation).

\section{Materials and Methods}\label{materials-and-methods}

\subsection{Data}\label{data}

We used both the HelixMTdb database \citep{Bolze2019} of publicly
available SNVs based on 195,983 individuals and the gnomAD database
\citep{Laricchia2022, Chen2023} of publicly available SNVs based on
56,406 individuals. In both databases, SNV frequencies were available at
each TLHG.

The paper underlying the HelixMTdb database \citep{Bolze2019} is not
published in a traditionally peer-reviewed journal, but there are
peer-reviews available at the preprint server, bioRxiv, hosting the
paper and associated TLHG and SNV data \citep{Bolze2019}. Following
personal communication with the authors, we have no reasons to doubt the
quality of the data.

We only considered homoplasmic positions where a variant had been
observed at least twice (globally, not within a TLHG). Insertions and
deletions (indels) were ignored, as these are prone to mutation,
sequencing errors, and alignment errors that lead to inconsistencies in
reported profiles.

\subsection{Inference of ``top-level''
haplogroup}\label{inference-of-top-level-haplogroup}

Haplogroup estimation with EMPOP/SAM2 \citep{Huber2018} is most reliable
when the entire mitogenome is available; the smaller the analysed
region, the less certain the estimation potentially becomes. As we only
needed the TLHG, we investigated if fewer positions were able to achieve
this goal, as this may be important in a forensic genetic context with
low-quality samples.

A total of 39 haplogroup-defining motifs were selected based on refined
haplogroup motifs. The source for these motifs was \citep{Dur2021}. The
selected 39 haplogroup motifs were: L0, L1, L5, L2, L6, L4, L3, M, M8,
C, Z, E, G, Q, D, N, I, W, Y, A, O, S, X, R, R9, B, HV, HV0, V, H, H2,
J, T, F, U, U8, U5, K, P. These haplogroups were selected to represent
all major lineages of human mitochondrial DNA (mtDNA) and largely
correspond to the haplogroups in HelixMTdb \citep{Bolze2019}, with minor
differences to preserve the precision of haplogroup estimation.

All positions defining these 39 haplogroup motifs were extracted,
resulting in a final dataset of 227 positions. The full list of these
227 positions is available in the supplementary material
(Table~\ref{tbl-227pos}).

We used two data sources for full mitogenomes. One was GenBank with
61,295 mitogenomes \citep{Huber2025}. The quality of these mitogenomes
was unknown and may even be varying. The GenBank sequences were
scrutinised and wrong/implausible mitotypes were removed
\citep{Huber2025}. For the remainder of this paper, when we refer to the
GenBank data, we refer to this scrutinised collection of sequences. The
other was high-quality published mitogenomes consisting of 1,327 U.S.
individuals (1,280 unique mitogenomes) \citep{Taylor2020}, 588 U.S.
individuals (556 unique mitogenomes), \citep{Just2015}, and 934 Swedish
individuals (837 unique mitogenomes) \citep{SturkAndreaggi2023}. Thus,
there was a total of 2,849 high-quality mitogenomes.

To obtain sufficiently large sample sizes and increase TLHG inference
accuracy, we collapsed TLHG L4, L5, and L6 into L4-6 and similarly, we
collapsed R and B into R/B.

For TLHG prediction, we both used the entire mitogenome and only the 227
selected positions (Table~\ref{tbl-227pos}). This left 17,197 different
227-position haplotypes for GenBank (of 61,295 originally) and 1,422
different 227-position haplotypes for the high-quality database (of
2,849 originally). SAM2 \citep{Huber2018} was used for the actual TLHG
prediction, both rank 1 and 2 were recorded.

\subsection{Mitogenome frequency estimation and weight of
evidence}\label{mitogenome-frequency-estimation-and-weight-of-evidence}

The positions of the mitogenome are statistically dependent, hence the
mitogenome frequency cannot simply be obtained by multiplying the
individual positions' frequencies
\citep{Andersen2021, Andersen2018mtDNA}.

Let \(X_1, X_2, \ldots, X_{p}\) be the state of the positions at the
mitogenome, e.g., relative to the Revised Cambridge Reference Sequence
(rCRS) \citep{Anderson1981, Andrews1999}. Thus, the frequency of the
mitogenome \(X = (X_1, X_2, \ldots, X_{p})\) is \begin{align}
P(X) = P(X_1, X_2, \ldots, X_{p}) .
\end{align} Standard telescoping from probability theory (factorisation
of simultanous probability by conditional probabilities) states the
general rule that \begin{align}
P(X_1, X_2, \ldots, X_{p}) = 
P(X_1) P(X_2, \ldots, X_{p} \mid X_1) =
P(X_1) P(X_2 \mid X_1) P(X_3, \ldots, X_{p} \mid X_1, X_2) 
\end{align} etc. If some positions can be known/shown to be
(conditionally) independent, then this expression can be simplified. For
example, if \(X_3, \ldots, X_{p}\) are conditionally independent of
\(X_1\) given \(X_2\), then
\(P(X_3, \ldots, X_{p} \mid X_1, X_2) = P(X_3, \ldots, X_{p} \mid X_2)\)
and \begin{align}
P(X_1) P(X_2 \mid X_1) P(X_3, \ldots, X_{p} \mid X_1, X_2) 
= 
P(X_1) P(X_2 \mid X_1) P(X_3, \ldots, X_{p} \mid X_2) .
\end{align} In this way, the long and complicated telescoping can be
simplified by (conditional) independence. This was previously done for Y
chromosomal short-tandem repeat (STR) markers
\citep{Andersen2018Y, Andersen2020Y}. In this paper we consider an even
simpler model that can be presented as follows.

Note that using fewer positions cannot result in a smaller mitogenome
frequency. In a simplified formulation, \begin{align}
P(X_1, X_2) \leq P(X_1) .
\end{align} In general, the probability of all positions matching is
less than (or equal to) the probability of only a subset matching.

Assume that \begin{align}
P(\text{TLHG} \mid X) 
= 1 ,
\end{align} i.e., we can assign the correct TLHG with certainty. Then it
follows that \begin{align} \label{eq:match-prob}
\begin{split}
P(\text{mitogenome}) 
&= P(X) \\
&= P(X) P(\text{TLHG} \mid X) \\
&= P(X, \text{TLHG}) \\
&= P(X_1, X_2, \ldots, X_{p}, \text{TLHG}) \\
&\leq P(X_k, \text{TLHG}) \\
&= P(\text{TLHG}) P(X_k \mid \text{TLHG})
\end{split}
\end{align} for any \(k\).

We can then calculate a weight of evidence as a likelihood ratio
(\(LR\)) by \begin{align}
  LR &= \frac{1}{P(\text{TLHG}) P(X_k \mid \text{TLHG})} \leq \frac{1}{P(X)} .
\end{align}

\subsection{Finer haplogrouping}\label{sec-finer-hg}

In the above, the TLHGs were used. This section presents the proof of an
important property under mild assumptions: If a finer subdivision (i.e.,
a more specific haplogrouping) was used instead and SNV frequency data
were available, then the estimated population frequency will be smaller
or equal to the one obtained using the coarser subdivision. In other
words, the \(LR\) will be higher when using a more specific
haplogrouping. Note that even if the SNV frequencies in the finer
subdivision are higher, this will get compensated for by a lower
subdivision frequency. Thus, by using the TLHG, we get an \(LR\) that is
equal to or lower than the \(LR\) that would be estimated from the
refined HG, if we had the refined HG frequency data.

To exemplify, assume that the TLHG is M. If the approach presented here
was used with the TLHG frequency and the within-TLHG SNV frequency to
obtain \(LR_{\text{M}}\), then, if we knew that the mitogenome was
actually from M1, and if we had SNV frequencies from M1, which we used
to obtain \(LR_{\text{M1}}\), then we are guaranteed, under mild
assumptions, that \(LR_{\text{M}} \leq LR_{\text{M1}}\).

Mathematically, the mtDNA haplogroup tree (the subdivisions) can be
described in many ways. Here, we assume that the subdivision structure
is a directed acyclic graph (DAG). Then, if \(X_k\) is the SNV chosen
and \(G\) (e.g., TLHG M) is an ancestor of \(H\) (e.g., HG M1), we show
that

\begin{align}
P(\text{mitogenome}) 
\leq P(H) P(X_k \mid H) 
\leq P(G) P(X_k \mid G) .
\end{align}

Thus, we can get closer to the true frequency by considering a
refinement of the TLHG, provided that relevant data are available.

This will first be shown for the estimator based on a single sample
(e.g., HelixMTdb \citep{Bolze2019}) used for both estimating the TLHG
frequency and the within-TLHG frequency. Afterwards, we give a numerical
example. Finally, we prove this fact in the general case.

\subsubsection{Estimator based on a single
sample}\label{estimator-based-on-a-single-sample}

Assume that we have a sample of \(n\) individuals of which \(n_G\)
belong to subdivision \(G\). Further, assume that \(m\) of the \(n\)
individuals have the SNV \(X_k\), and that \(m_G\) of the individuals in
\(G\) have the SNV \(X_k\). Then \(P(G) P(X_k \mid G)\) is estimated by
\begin{align}
\frac{n_G}{n} \cdot \frac{m_G}{n_G} = 
\frac{m_G}{n} ,
\end{align} i.e., the number of individuals in subdivision \(G\) that
have SNV \(X_k\) out of all \(n\) individuals.

Now, if \(G\) is an ancestor of \(H\) in our subdivisions represented by
a DAG, it follows that \(m_H \leq m_G\) and hence \begin{align}
\frac{m_H}{n} \leq \frac{m_G}{n} .
\end{align}

This means that the \(LR\)s are \begin{align}
LR_H = \frac{n}{m_H} \geq \frac{n}{m_G} = LR_G .
\end{align}

Thus, the \(LR\) based on the subdivision H cannot be smaller than the
\(LR\) obtained by using the subdivision G.

\subsubsection{A numerical example}\label{a-numerical-example}

\begin{figure}

\centering{

\includegraphics[width=0.6\linewidth,height=\textheight,keepaspectratio]{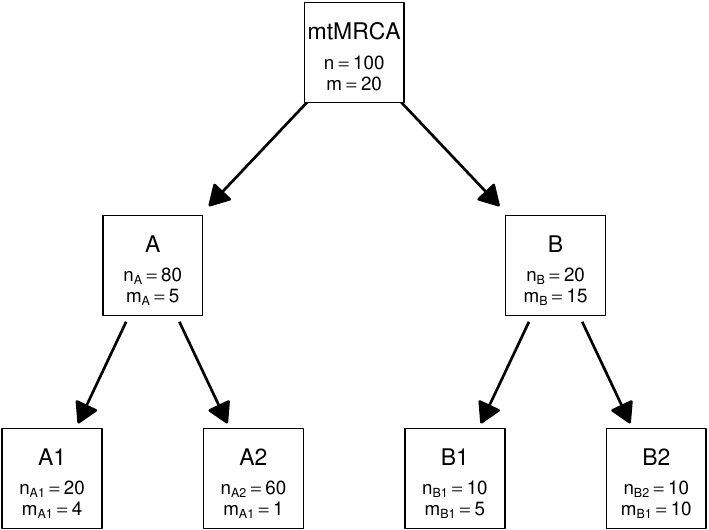}

}

\caption{\label{fig-LR-subdivision}An example of subdivisions' effect on
the LR.}

\end{figure}%

Consider the case illustrated by Figure~\ref{fig-LR-subdivision}. Here,
using TLHG A and the SNV \(X_k\), then

\begin{align}
P(\text{A}) P(X_k \mid \text{A}) = \frac{n_A}{n} \cdot \frac{m_A}{n_A} =
\frac{80}{100} \cdot \frac{5}{80} = 
\frac{5}{100}, 
\end{align} and using a refinement with A1 instead yields \begin{align}
P(\text{A1}) P(X_k \mid \text{A1}) = \frac{n_{A1}}{n} \cdot \frac{m_{A1}}{n_{A1}} =
\frac{20}{100} \cdot \frac{4}{20} = 
\frac{4}{100} .
\end{align}

We can then calculate \begin{align}
LR_A &= \frac{1}{P(\text{A}) P(X_k \mid \text{A})} = \frac{100}{5} = 20 \quad \text{and} \\
LR_{A1} &= \frac{1}{P(\text{A1}) P(X_k \mid \text{A1})} = \frac{100}{4} = 25.
\end{align}

Note that \(LR_A \leq LR_{A1}\). Here, SNV \(X_k\) is a lot more common
in A1 than in A (\(\frac{4}{20} = 0.2 > \frac{5}{80} = 0.0625\)). Thus,
any potential increased SNV frequency is balanced out by a decreased
subdivision frequency.

\subsubsection{General proof}\label{general-proof}

Let \(\Omega_1\) be a coarse (e.g.~TLHG) subdivision, i.e.~partitioning,
and let \(\Omega_2\) be a finer subdivision. Let \(G \in \Omega_1\) be
the subdivision for the mitogenome, \(X\), in question and let
\(H \in \Omega_2\) such that \(H \subseteq G\). In DAG terms, the node
representing \(G\) is an ancestor of the node representing \(H\).

We assume that \(P(H \mid X) = 1\), i.e., we can determine the correct
subdivision given the mitogenome, \(X\), with certainty for the
subdivision level chosen. As \(H \subseteq G\), this implies that
\(P(G \mid X) = 1\). We want to show that \begin{align}
P(\text{mitogenome}) 
\leq P(H) P(X_k \mid H)
\leq P(G) P(X_k \mid G) .
\end{align} Thus, using a finer subdivision, we get close to
\(P(\text{mitogenome})\) from above. We have the first inequality from
earlier in Eq. \eqref{eq:match-prob}. Thus, we need to show that
\begin{align}
P(H) P(X_k \mid H) \leq P(G) P(X_k \mid G) .
\end{align}

By monotonicity of probabilities, \(P(H) \leq P(G)\) as
\(H \subseteq G\). The same holds for joint probabilities, giving
\begin{align}
P(H) P(X_k \mid H) 
= P(H, X_k)
\leq P(G, X_k) 
= P(G) P(X_k \mid G) .
\end{align}

\subsection{SNV frequencies}\label{snv-frequencies}

Reliable estimates of the SNV frequencies within a TLHG,
\(P(X_k \mid \text{TLHG})\), are required, with explicit consideration
of the geographic location of interest. Here we use the HelixMTdb
database \citep{Bolze2019} or the gnomAD database
\citep{Laricchia2022, Chen2023}.

Since SNV frequencies for refined HG data were unavailable, we compared
the HelixMTdb (\(n =\) 195,983) SNV frequencies with those from gnomAD
(\(n =\) 56,406) \citep{Laricchia2022, Chen2023} for each position and
TLHG to assess consistency between the respective TLHG-derived SNV
frequencies of the two databases.

If a SNV was observed \(x_{\text{Helix}}\) times in HelixMTdb out of
\(n_{\text{Helix}}\) individuals in the TLHG in question, then the SNV
frequency, \(P(X_k \mid \text{TLHG})\), was estimated by \begin{align}
\hat{p}_{\text{Helix}} 
&= \frac{x_{\text{Helix}}}{n_{\text{Helix}}} ,
\end{align} and similarly for the gnomAD data.

Other estimators could be used, e.g., Bayesian ones, as explained for
Y-STR haplotypes by \citep{Andersen2017Y}.

If we have data from multiple databases, e.g., both HelixMTdb and
gnomAD, then we can make a pooled estimate of the SNV frequency,
\(P(X_k \mid \text{TLHG})\), by \begin{align}
\hat{p}_{\text{pooled}} 
&= \frac{x_{\text{Helix}} + x_{\text{gnomAD}}}{n_{\text{Helix}} + n_{\text{gnomAD}}} .
\end{align} Note that the databases' estimates were weighted according
to their sizes.

We only used variants where the reported bases and variants were a
single base. We also only used SNVs that were observed in the TLHG in
question. Potentially, this could be extended, e.g.~by exploiting if a
SNV had been seen in other TLHGs.

\subsection{Software implementation}\label{software-implementation}

We have developed an open-source R package \texttt{mitofreq} that
implements our method and includes the HelixMTdb and gnomAD data (both
SNV frequencies and TLHG distributions). The development version is
available at \url{https://github.com/mikldk/mitofreq}. We have also made
a development version of a Shiny webapp (included in the R package)
available for live demonstration purposes at
\url{https://mikldk.shinyapps.io/MitoFREQ/}.

The HelixMTdb database \citep{Bolze2019} had information on 10,253
positions out of the 16,569 mitogenome positions, and 8,327 positions
were left after the filtering. Refer to the R package
\texttt{mitofreq}'s dataset \texttt{d\_helix}.

The gnomAD database \citep{Laricchia2022, Chen2023} had information on
12,749 positions out of the 16,569 mitogenome positions, and 7,223
positions were left after the filtering. Refer to the R package
\texttt{mitofreq}'s dataset \texttt{d\_gnomAD}.

\section{Results}\label{results}

\subsection{Inference of TLHG}\label{inference-of-tlhg}

We inferred the TLHG based on the entire mitogenome using EMPOP/SAM2
\citep{Huber2018}. This was compared to the results using only the 227
positions where we recorded both the most likely TLHG (rank 1) and the
second most likely (rank 2) TLHG predicted by SAM2.

The rank 1 TLHG prediction results are presented in
Table~\ref{tbl-TLHG-rank1}. If considering both rank 1 and rank 2, the
TLHG prediction results are presented in Table~\ref{tbl-TLHG-rank12}.

\begin{table}

\caption{\label{tbl-TLHG-rank1}Rank 1 TLHG prediction results.}

\centering{

\begin{tabular}[t]{lrrr}
\toprule
Dataset & Concordant & Discordant & Concordance [\%]\\
\midrule
GenBank & 60,698 & 597 & 99.0\\
SWE & 929 & 5 & 99.5\\
US2015 & 587 & 1 & 99.8\\
US2020 & 1,323 & 4 & 99.7\\
\bottomrule
\end{tabular}

}

\end{table}%

\begin{table}

\caption{\label{tbl-TLHG-rank12}Rank 1 or 2 TLHG prediction results.}

\centering{

\begin{tabular}[t]{lrrr}
\toprule
Dataset & Concordant & Discordant & Concordance [\%]\\
\midrule
GenBank & 61,233 & 62 & 99.9\\
SWE & 934 & 0 & 100.0\\
US2015 & 588 & 0 & 100.0\\
US2020 & 1,327 & 0 & 100.0\\
\bottomrule
\end{tabular}

}

\end{table}%

\subsection{Single nucleotide
variants}\label{single-nucleotide-variants}

\begin{figure}

\centering{

\includegraphics[width=0.5\linewidth,height=\textheight,keepaspectratio]{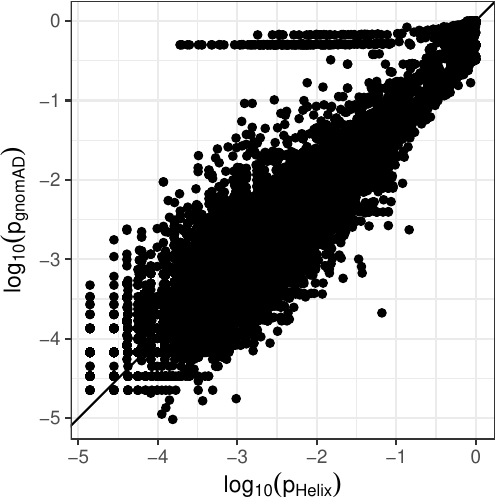}

}

\caption{\label{fig-SNV-frequency-Helix-gnomAD}Comparison of the SNV
frequencies, \(p_{\text{Helix}}\) and \(p_{\text{gnomAD}}\). Each point
represents an alternative allele at a mitogenome position in a TLHG. The
Pearson correlation of the \(\log_{10}\) transformed SNV frequencies was
0.982. We only considered SNVs observed in both databases.}

\end{figure}%

The SNV frequencies are compared in
Figure~\ref{fig-SNV-frequency-Helix-gnomAD}. Each point represents a
single SNV in a single TLHG. Thus, the frequencies in Helix and gnomAD
are comparable. We only considered SNVs where alternative variants were
observed in both databases (4,762 SNVs).

We found a large Pearson correlation of the \(\log_{10}\) transformed
SNV frequencies of 0.982. Note that the increased variability for rare
SNVs is not surprising, as really large databases are needed for
estimating rare events; that the database sizes are totals; and the SNV
frequencies are estimated within each TLHG. Also note that common SNVs
in the gnomAD database had a wide range of frequencies in the Helix
database (see the horizontal line of points at the top), indicating that
recurring errors (or under-sampling) may be present in the gnomAD
database.

\subsection{Weight of evidence}\label{weight-of-evidence}

\begin{table}

\caption{\label{tbl-Brenner-GGT-LR}\(LR\) values for singletons by
Brenner's \(\kappa\) estimator and Cereda's estimator based on the
Generalized-Good-Turing estimator (CGGT). Note that the entire variant
strings were used to determine singleton and doubleton proportions.
Refer to the text for a discussion.}

\centering{

\centering\begingroup\fontsize{8}{10}\selectfont

\begin{tabular}[t]{lrrrrrrr}
\toprule
Dataset & Database size (n) & Singletons (s) & s/n [\%] & Doubletons (d) & d/n [\%] & Brenner's LR & CGGT LR\\
\midrule
GenBank & 61,295 & 42,614 & 69.5 & 3,466 & 5.7 & 201,124 & 376,807\\
SWE & 934 & 778 & 83.3 & 47 & 5.0 & 5,604 & 7,730\\
US2015 & 588 & 573 & 97.4 & 6 & 1.0 & 23,128 & 28,077\\
US2020 & 1,327 & 1,265 & 95.3 & 20 & 1.5 & 28,445 & 41,966\\
\bottomrule
\end{tabular}
\endgroup{}

}

\end{table}%

As reference, we calculated \(LR\) values for singletons with the count
estimators Brenner's \(\kappa\) estimator \citep{Brenner2010} and
Cereda's estimator based on the Generalized-Good-Turing (CGGT) estimator
\citep{Good1953, Cereda2016}. The results are shown in
Table~\ref{tbl-Brenner-GGT-LR}. Note that in our study, we calculated
the singleton and doubleton proportions from the entire variant string,
where no variants were removed (i.e., SNVs within poly-stretches at
303-315, 16180-16194, and 513-525 were not removed, although indels
within these regions were ignored). If variants were removed, this would
lead to lower singleton and doubleton proportions and thereby lower
\(LR\)s. As an example, \citep{Just2015} reports a maximal Brenner
\(LR\) of 14,450 (for an African American dataset with 170 samples)
whereas our simple method gives a Brenner \(LR\) of 23,128.

\begin{figure}

\centering{

\includegraphics[width=0.75\linewidth,height=\textheight,keepaspectratio]{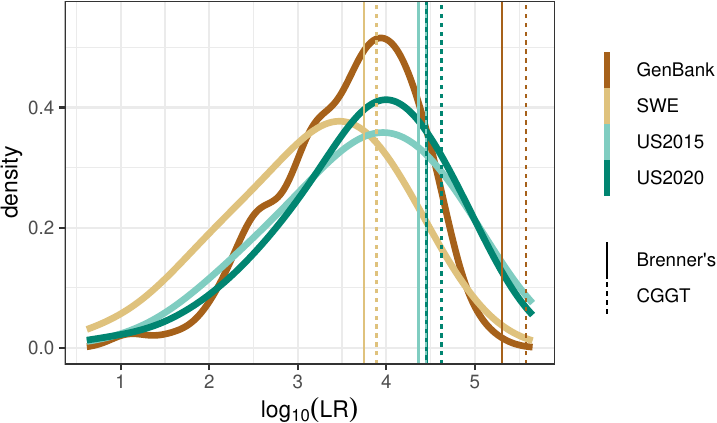}

}

\caption{\label{fig-LR}Distribution of \(\log_{10}(LR)\) values using
the smallest one from the rank 1 and rank 2 TLHG prediction and pooled
SNV frequencies. The reference \(\log_{10}(LR)\)s for singletons are
from the count estimators in Table~\ref{tbl-Brenner-GGT-LR}.}

\end{figure}%

\begin{figure}

\centering{

\includegraphics[width=1\linewidth,height=\textheight,keepaspectratio]{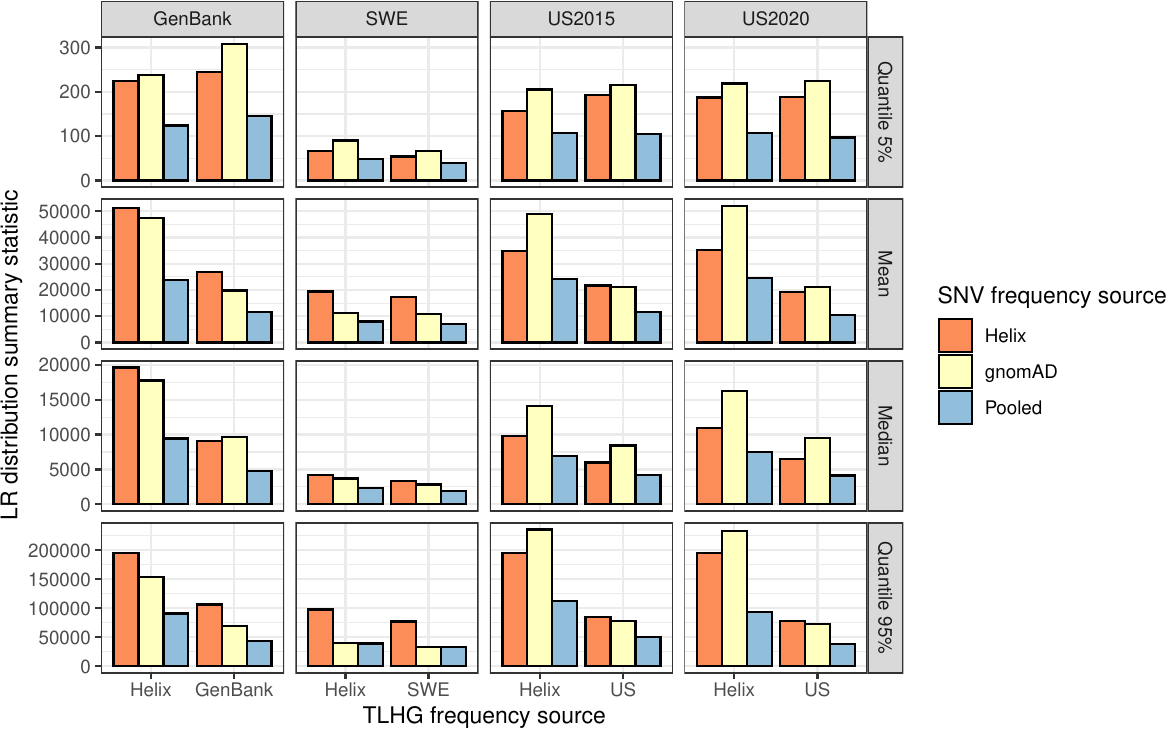}

}

\caption{\label{fig-LR-dist}Summary statistics of \(LR\) distributions
using the smallest one from the rank 1 and rank 2 TLHG prediction. The
columns correspond to the datasets, and the rows correspond to the
summary statistics.}

\end{figure}%

\(LR\)s were calculated by combining the rarest SNV frequency within a
TLHG with the TLHG frequency. We used the smallest \(LR\) obtained
between the rank 1 and rank 2 TLHG prediction.

The distributions of the \(\log_{10} (LR)\) values are shown in
Figure~\ref{fig-LR} for pooled SNV frequencies and TLHG frequencies as
follows: GenBank use GenBank, US2015 and US2020 use Helix, and SWE use
SWE to infer TLHG frequencies based on the largest and most appropriate
dataset for the population.

In Figure~\ref{fig-LR-dist}, summary statistics of the \(LR\)
distributions are given for different combinations of SNV and TLHG
frequency sources.

\subsection{Examples with frequent and rare
haplotypes}\label{examples-with-frequent-and-rare-haplotypes}

Here we demonstrate our method on a number of examples based on rare and
frequent haplotypes in the GenBank data \citep{Huber2025} with 61,295
mitogenomes in larger and smaller clades. The chosen haplotypes'
haplogroups and variants can be seen in
Table~\ref{tbl-GenBank-examples-haplotypes} in Supplementary Material.
The results (chosen SNVs with minimal frequency within the TLHGs and
\(LR\)s) are shown in Table~\ref{tbl-GenBank-examples-LRs}.

\begin{table}

\caption{\label{tbl-GenBank-examples-LRs}MitoFREQ results for the
selected examples from GenBank. The haplogroups and haplotypes can be
seen in Table~\ref{tbl-GenBank-examples-haplotypes}. Helix, gnomAD, and
Pooled refers to the SNV selected with the lowest SNV frequency within
the TLHG. The \(LR\)s are based on Helix TLHG frequencies.}

\centering{

\centering\begingroup\fontsize{8}{10}\selectfont

\begin{tabular}[t]{lllrrrrrrr}
\toprule
\multicolumn{4}{c}{ } & \multicolumn{3}{c}{Min. frequency SNV} & \multicolumn{3}{c}{$LR$} \\
\cmidrule(l{3pt}r{3pt}){5-7} \cmidrule(l{3pt}r{3pt}){8-10}
Label & Region & TLHG & Count & Helix & gnomAD & Pooled & Helix & gnomAD & Pooled\\
\midrule
Large: Frequent & Africa & L0 & 14 & 7154G & 597T & 7154G & 195,983 & 624,696 & 169,444\\
Large: Frequent & Asia & M & 72 & 417A & 417A & 417A & 19,598 & 5,688 & 9,447\\
Large: Frequent & Europe & H & 68 & 14902T & 14902T & 14902T & 224 & 280 & 232\\
\addlinespace
Large: Rare & Africa & L0 & 1 & 13928C & 8994A & 16214T & 195,983 & 624,696 & 91,187\\
Large: Rare & Asia & M & 1 & 8545A & 15412C & 15412C & 195,983 & 153,568 & 69,910\\
Large: Rare & Europe & H & 1 & 225A & 13101C & 4820A & 13,999 & 2,619 & 1,939\\
\addlinespace
Small: Frequent & Africa & L4-6 & 2 & 15924G & 16166G & 7521A & 195,983 & 1,131,356 & 189,302\\
Small: Frequent & Asia & M & 4 & 89C & 12940A & 16343G & 195,983 & 153,509 & 106,569\\
Small: Frequent & Europe & I & 4 & 8562T & 5773A & 5096C & 195,983 & 124,337 & 7,969\\
\addlinespace
Small: Rare & Africa & L4-6 & 1 & 8152A & 16166G & 8152A & 195,983 & 1,131,356 & 380,831\\
Small: Rare & Asia & M & 1 & 228T & 16286T & 10598G & 195,983 & 230,234 & 90,922\\
Small: Rare & Europe & I & 1 & 150T & 6227C & 6227C & 2,000 & 1,974 & 1,562\\
\bottomrule
\end{tabular}
\endgroup{}

}

\end{table}%

\section{Discussion}\label{discussion}

In this paper, we have presented a method for evaluating the evidential
strength of mitogenomes that does not use the entire mitogenome, but
only the TLHG and one (rare) SNV. This is especially important in
forensic genetics where samples can be heavily degraded with partial
profiles, and it can be impossible to infer the entire mitogenome
sequence. With our method, the most rare SNV with a confirmed match in
the database can be used.

Our method is based on estimating mitogenome frequencies. As discussed
by \citep{Andersen2021}, using population frequencies as match
probabilities requires a low profile mutation rate. There has also been
a discussion on how to report lineage marker evidence, especially Y-STR
\citep{Andersen2017Y, Andersen2021, Roewer2020, Bright2024}. We propose
to follow \citep{Andersen2023, Bright2024} with the propositions

\begin{itemize}
\tightlist
\item
  Proposition 1: The source of the mtDNA is the person of interest.
\item
  Proposition 2: The source of the mtDNA is a random individual from the
  population.
\end{itemize}

including, in the body of the statement, the fact that close matrilineal
relatives stand an enhanced chance of sharing the mitogenome, that the
weight of evidence calculations are not valid for these close
matrilineal relatives, and the random individual in Proposition 2 is
another individual than the person of interest.

An important property of our method is that finer haplogrouping results
in population frequency estimates which are smaller than or equal to
those obtained when using a coarser haplogrouping, and thus provide
\(LR\)s larger than or equal to those obtained from the coarser
haplogrouping as described in Section~\ref{sec-finer-hg}.

Reliable estimates of the SNV frequencies within a TLHG,
\(P(X_k \mid \text{TLHG})\), are required, with explicit consideration
of the geographic location of interest.

In Figure~\ref{fig-SNV-frequency-Helix-gnomAD} we compared SNV frequency
estimates from HelixMTdb \citep{Bolze2019} and gnomAD
\citep{Laricchia2022, Chen2023}. We found a large correlation (0.982).
Note that the increased variability for rare SNVs is not surprising, as
really large databases are needed for estimating rare events, and that
the database sizes are totals, and the SNV frequencies are estimated
within each TLHG.

The results in Table~\ref{tbl-GenBank-examples-LRs} indicate that
mitogenome frequency estimates and consequently the resulting \(LR\)s
vary across continental regions. European lineages appear to be better
represented in current population resources, whereas African lineages
remain comparatively undersampled. This imbalance highlights the
importance of considering the continental context when interpreting
mtDNA evidence, as database composition can influence the inferred
evidential value.

These analyses also rely on the assumption that the probability of a SNV
within a given TLHG is independent of geographic origin. This is
potentially an oversimplification, as haplogroup-dependent SNVs may
differ between populations. Still, HelixMTdb and gnomAD show high
correlation (0.982, cf. Figure~\ref{fig-SNV-frequency-Helix-gnomAD}) in
SNV frequencies across regions, suggesting that overlapping SNVs are
relatively robust to such effects. Restricting \(LR\) calculations to
these shared variants and ensuring population-appropriate reference data
can help avoid overestimating the rarity of mitogenomes common in
undersampled areas.

As only TLHG were needed, the 227 positions that we found were
sufficient, especially when we considered both rank 1 and 2 of the TLHG
prediction (Table~\ref{tbl-TLHG-rank1} and Table~\ref{tbl-TLHG-rank12}).

In this study, we used TLHG distributions from the available datasets,
but in practise any TLHG distribution can be used together with the SNV
frequencies. As an example, a country-specific TLHG distribution can be
used together with a pooled SNV frequency estimate from Helix and
gnomAD.

We compared our method to counting estimators (Brenner's \(\kappa\)
estimator \citep{Brenner2010} and CGGT estimator
\citep{Good1953, Cereda2016}) that do not exploit genetic information
and require that the range of the profile in question aligns with that
of the database (e.g.~entire mitogenome). We found that the counting
estimators corresponded approximately to the average \(\log_{10}(LR)\)
obtained from our method applied on US and SWE databases
(Figure~\ref{fig-LR}). This was not true for the Genbank data. This
could potentially be caused by a difference in homogeneity of the
databases.

Our proposed model has an important potential extension: More SNVs could
in principle be used if data was available - either in a model that
takes dependency into account, or by ensuring conditional independence
between the SNVs within TLHG. More work on this is needed, and can
potentially use graphical models similar to what have been done for
Y-STR profiles \citep{Andersen2018Y, Andersen2020Y}. It can also mean
that TLHGs are not needed at all, which would potentially make the model
more robust. The downside is that it requires many mitogenome profiles
(and not just within-TLHG SNV frequencies), which we currently only have
available via GenBank, and the quality of those profiles for accurate,
forensic match statistics is unknown.

In our comparisons, we did not exclude any positions before calculating
the singleton and doubleton proportions. In general, we recommend: 1)
defining the positions to ignore (related to, e.g., poly-stretches,
heteroplasmies, diseases etc.) in guidelines, 2) using only the filtered
mitogenomes in both comparisons between a trace profile and a profile
from a person of interest, and 3) including only the filtered profiles
in the database for frequency estimation. This would mean that, e.g.,
singleton proportions, would be calculated based on the filtered
variants.

\section{Disclaimer}\label{disclaimer}

The assertions herein are those of the authors and do not necessarily
represent the official position of the United States Department of
Defense, the Defense Health Agency, or its entities including the Armed
Forces Medical Examiner System.

\section*{References}\label{references}
\addcontentsline{toc}{section}{References}

\FloatBarrier
\clearpage

\newpage

\section{Supplementary material}\label{supplementary-material}

\renewcommand{\thefigure}{S\arabic{figure}}
\setcounter{figure}{0}
\renewcommand{\thetable}{S\arabic{table}}
\setcounter{table}{0}

\begin{table}

\caption{\label{tbl-227pos}227 positions used for ``top-level''
haplogroups (TLHG) prediction.}

\centering{

\begin{tabular}[t]{llllllll}
\toprule
Position & Position & Position & Position & Position & Position & Position & Position\\
\midrule
72 & 1,018 & 4,529 & 7,256 & 10,115 & 12,612 & 14,783 & 16,187\\
73 & 1,048 & 4,580 & 7,389 & 10,238 & 12,705 & 14,791 & 16,189\\
146 & 1,243 & 4,715 & 7,521 & 10,310 & 12,720 & 14,798 & 16,192\\
150 & 1,438 & 4,769 & 7,598 & 10,398 & 12,771 & 14,905 & 16,213\\
152 & 1,461 & 4,824 & 7,972 & 10,400 & 12,940 & 14,959 & 16,223\\
\addlinespace
182 & 1,719 & 4,833 & 8,206 & 10,463 & 12,950 & 15,043 & 16,224\\
185 & 1,736 & 4,883 & 8,251 & 10,550 & 13,105 & 15,244 & 16,230\\
189 & 1,811 & 4,917 & 8,392 & 10,589 & 13,263 & 15,289 & 16,231\\
195 & 1,888 & 4,964 & 8,404 & 10,664 & 13,276 & 15,301 & 16,241\\
199 & 2,416 & 5,046 & 8,468 & 10,688 & 13,368 & 15,326 & 16,260\\
\addlinespace
204 & 2,706 & 5,108 & 8,584 & 10,810 & 13,500 & 15,452 & 16,270\\
207 & 2,758 & 5,178 & 8,655 & 10,873 & 13,506 & 15,487 & 16,278\\
235 & 2,885 & 5,267 & 8,697 & 10,915 & 13,590 & 15,499 & 16,290\\
247 & 3,027 & 5,417 & 8,701 & 10,978 & 13,626 & 15,607 & 16,292\\
249 & 3,423 & 5,442 & 8,790 & 11,116 & 13,650 & 15,784 & 16,294\\
\addlinespace
250 & 3,480 & 5,460 & 8,794 & 11,251 & 13,708 & 15,884 & 16,298\\
263 & 3,505 & 5,843 & 8,860 & 11,299 & 13,710 & 15,904 & 16,304\\
295 & 3,516 & 6,002 & 8,994 & 11,467 & 13,780 & 15,924 & 16,311\\
315 & 3,552 & 6,185 & 9,042 & 11,674 & 13,789 & 15,928 & 16,319\\
459 & 3,594 & 6,221 & 9,055 & 11,719 & 13,928 & 16,048 & 16,327\\
\addlinespace
489 & 3,666 & 6,284 & 9,090 & 11,743 & 13,966 & 16,069 & 16,362\\
573 & 3,705 & 6,371 & 9,140 & 11,914 & 14,167 & 16,126 & 16,390\\
663 & 3,970 & 6,392 & 9,221 & 11,947 & 14,178 & 16,129 & 16,391\\
709 & 4,104 & 6,752 & 9,332 & 12,007 & 14,318 & 16,148 & 16,519\\
750 & 4,117 & 6,755 & 9,347 & 12,308 & 14,470 & 16,166 & \\
\addlinespace
769 & 4,216 & 7,028 & 9,540 & 12,372 & 14,560 & 16,181 & \\
770 & 4,248 & 7,055 & 9,545 & 12,414 & 14,569 & 16,182 & \\
825 & 4,312 & 7,146 & 9,698 & 12,432 & 14,693 & 16,183 & \\
961 & 4,491 & 7,196 & 10,034 & 12,501 & 14,766 & 16,185 & \\
\bottomrule
\end{tabular}

}

\end{table}%

\begingroup\fontsize{8}{10}\selectfont

\begin{longtable}[t]{llllr>{\raggedright\arraybackslash}p{8cm}}

\caption{\label{tbl-GenBank-examples-haplotypes}Haplotypes for the
selected examples from GenBank. Count refers to the count in the GenBank
dataset. TLHG refers to the TLHG used in MitoFREQ calculations.}

\tabularnewline

\toprule
Label & Region & Haplogroup & TLHG & Count & Haplotype\\
\midrule
Large: Frequent & Africa & L0d2a1a & L0 & 14 & 73G 146C 152C 195C 198T 247A 309.1C 315.1C 498del 523del 524del 597T 750G 769A 825A 1018A 1048T 2706G 2758A 2885C 3516A 3594T 3756G 3981G 4025T 4044G 4104G 4225G 4232C 4312T 4769G 5153G 5442C 6185C 6815C 7028T 7146G 7154G 7256T 7521A 8113A 8152A 8251A 8392A 8468T 8545A 8655T 8701G 8860G 9042T 9347G 9540C 9755A 10398G 10589A 10664T 10688A 10810C 10873C 10915C 11719A 11854C 11914A 12007A 12121C 12172G 12234G 12705T 12720G 12810G 13105G 13276G 13506T 13650T 14221C 14766T 15326G 15466A 15766G 15930A 15941C 16129A 16187T 16189C 16223T 16230G 16243C 16311C 16390A 16519C\\
Large: Frequent & Asia & M23 & M & 72 & 73G 152C 195C 204C 263G 315.1C 417A 489C 533G 750G 1438G 4769G 7028T 8188G 8360G 8701G 8860G 9438A 9540C 9545G 10142T 10295G 10398G 10400T 10873C 11569C 11719A 11899C 12279G 12618A 12705T 14766T 14783C 15025T 15043A 15301A 15326G 16223T 16263C 16311C 16519C\\
Large: Frequent & Europe & H1e1a & H & 68 & 263G 750G 1438G 3010A 4769G 5460A 8512G 8860G 14902T 15326G 16519C\\
\addlinespace
Large: Rare & Africa & L0k1a1d & L0 & 1 & 73G 146C 152C 189G 195C 198T 207A 247A 309del 315.1C 523del 524del 750G 769A 825A 850C 1018A 1048T 1243C 1438G 2706G 2758A 2836A 2885C 3516A 3594T 4104G 4312T 4541A 4586C 4769G 4907C 5442C 5582G 5811G 6185C 6938T 7028T 7146G 7256T 7257G 7521A 8468T 8655T 8701G 8860G 8911C 8994A 9042T 9136G 9347G 9540C 9731T 9818T 10398G 10499G 10589A 10664T 10688A 10810C 10873C 10876G 10915C 10920T 10939T 11296T 11299C 11653G 11719A 11914A 12007A 12070A 12705T 13020C 13105G 13276G 13506T 13590A 13650T 13819C 13928C 14020C 14182C 14371C 14374C 14766T 15326G 16166C 16172C 16187T 16189C 16209C 16214T 16223T 16230G 16278T 16291A 16311C 16519C\\
Large: Rare & Asia & M7b2 & M & 1 & 73G 263G 309.1C 315.1C 489C 523del 524del 723G 750G 1438G 2706G 3140G 4071T 4769G 5324T 6216C 6455T 7028T 8027A 8545A 8701G 8860G 9039A 9540C 9824C 10398G 10400T 10873C 11347G 11719A 12352G 12405T 12705T 14284T 14552G 14766T 14783C 14798C 15043A 15301A 15326G 15412C 15942C 16184T 16223T 16311C\\
Large: Rare & Europe & H8 & H & 1 & 146C 225A 263G 315.1C 709A 750G 1438G 4769G 4820A 8860G 13101C 15326G 16288C 16362C\\
\addlinespace
Small: Frequent & Africa & L7a & L4-6 & 2 & 73G 146C 152C 182T 195C 198T 247A 263G 315.1C 550G 750G 769A 825A 1018A 1438G 2706G 3334G 3345C 3423C 3594T 3918A 4092A 4104G 4679C 4769G 6164T 6260A 6527G 7028T 7256T 7521A 7765G 8450C 8485A 8655T 8701G 8860G 9540C 9947A 10398G 10688A 10775A 10810C 10873C 10920T 11023G 11719A 11734G 11809C 12021C 12432T 12669T 12705T 12892C 13506T 13581C 13650T 14020C 14548G 14766T 15236G 15301A 15326G 15924G 15941C 16042A 16129A 16145A 16166G 16187T 16189C 16223T 16278T 16311C 16362C\\
Small: Frequent & Asia & Q1a1a & M & 4 & 73G 89C 92A 146C 195C 208C 263G 309.1C 315.1C 489C 750G 1438G 2706G 4117C 4703C 4769G 5112A 5460A 5843G 7028T 7681T 8701G 8790A 8860G 8964T 9266A 9540C 10398G 10400T 10873C 11719A 12705T 12940A 13086T 13500C 14025C 14766T 14783C 15043A 15301A 15326G 16129A 16144C 16148T 16222T 16241G 16265C 16311C 16343G 16497G\\
Small: Frequent & Europe & I5a2a & I & 4 & 73G 199C 204C 250C 263G 315.1C 573.1C 750G 1438G 1719A 2706G 3615G 3705A 4529T 4769G 5074C 5096C 5773A 7028T 8251A 8562T 8742G 8860G 10034C 10238C 10398G 11719A 12501A 12705T 13780G 14233G 14766T 15043A 15326G 15924G 15941C 16086C 16129A 16148T 16223T 16391A 16519C\\
\addlinespace
Small: Rare & Africa & L7b2a1 & L4-6 & 1 & 73G 152C 182T 247A 263G 315.1C 523del 524del 593C 750G 769A 825A 1018A 1438G 2706G 3204T 3423C 3592A 3594T 3705A 4104G 4769G 5237A 5530T 6044C 6527G 7028T 7256T 7389C 7521A 7711C 8059T 8152A 8176C 8270T 8281del 8282del 8283del 8284del 8285del 8286del 8287del 8288del 8289del 8634C 8655T 8701G 8830T 8860G 8992T 9512T 9540C 9755A 10398G 10688A 10810C 10873C 11719A 11809C 12178T 12432T 12651C 12681C 12705T 13506T 13650T 14348C 14766T 15326G 15911G 15927A 16166G 16172C 16187T 16189C 16209C 16213A 16223T 16278T 16311C\\
Small: Rare & Asia & Q2b & M & 1 & 73G 152C 228T 263G 309.1C 315.1C 489C 523del 524del 744G 750G 1438G 1462A 2706G 4117C 4769G 5456T 5843G 7028T 8701G 8790A 8860G 9540C 10398G 10400T 10598G 10873C 11061T 11719A 12267G 12268C 12705T 12940A 13500C 14766T 14783C 15043A 15070T 15301A 15326G 16066G 16129A 16145A 16223T 16286T\\
Small: Rare & Europe & I1b & I & 1 & 73G 150T 199C 204C 250C 263G 315.1C 455.1T 573.1C 573.2C 573.3C 573.4C 573.5C 750G 1438G 1719A 2706G 3777C 4529T 4769G 6227C 6734A 7028T 8251A 8860G 9966A 10034C 10238C 10398G 11719A 12501A 12705T 13780G 14766T 15043A 15045A 15326G 15924G 16129A 16223T 16311C 16391A 16519C\\
\bottomrule

\end{longtable}

\endgroup{}

  \bibliography{bibliography.bib}

\end{document}